\documentclass[12pt]{article}
\usepackage{amssymb}
\usepackage{amsxtra}
\usepackage{amsmath}
\usepackage{amstext}
\usepackage{amsthm}
\usepackage{amsbsy}
\usepackage{latexsym}
\usepackage{amscd}
\usepackage{eucal}
\usepackage[dvips]{graphicx}
\usepackage{graphics}
\usepackage{float}
\usepackage{anysize}
\usepackage{xcolor}
\usepackage[font=small,labelfont=bf]{caption}

\usepackage{authblk}

\title{The role of interparticle interaction and environmental coupling in a two-particle open quantum system}
\author[a,b]{Humberto G. Laguna}
\author[c]{Robin P. Sagar}
\author[d]{David G. Tempel}
\author[d]{Al{\'a}n Aspuru-Guzik}

\affil[a]{Departamento de Matem{\'a}ticas, Facultad de Ciencias, Universidad Nacional Aut{\'o}noma de M{\'e}xico, Circuito Exterior, Ciudad Universitaria, D.F. 04510, M{\'e}xico.}
\affil[b]{Centro de Ciencias de la Complejidad, Universidad Nacional Aut{\'o}noma de M{\'e}xico, Circuito Exterior, Ciudad Universitaria, D.F. 04510, M{\'e}xico.}
\affil[c]{Departamento de Qu{\'i}mica, Universidad Aut{\'o}noma Metropolitana, San Rafael Atlixco No. 186, Iztapalapa, 09340, M{\'e}xico D.F., M{\'e}xico.}
\affil[d]{Department of Chemistry and Chemical Biology, Harvard University, 12 Oxford Street, Cambridge, MA 02138, United States.}

\date{\today}

\begin{document}
\maketitle

\begin{abstract}

The effects of bath coupling on an interacting two-particle quantum system are studied using tools from information theory.
Shannon entropies of the one (reduced) and two-particle distribution functions in position, momentum and separable phase-space are examined.
Results show that the presence of the bath leads to a delocalization of the distribution functions in position space, and a localization in momentum space.
This can be interpreted as a loss of information in position space and a gain of information in momentum space. 
The entropy sum of the system, in the presence of a bath, is shown to be dependent on the strength of the interparticle potential and also on the strength of the
coupling to the bath. The statistical correlation between the particles,
and its dependence on the bath and interparticle potential, is examined using mutual information. A stronger repulsive potential between particles, in the presence of the bath, yields a smaller correlation between the particles positions, and a larger one between their momenta.

\end{abstract}

\newpage

\section{Introduction}

The study of open quantum systems (OQS) gives insights into a variety of theoretical and technological problems, such as the eventual construction of a quantum computer \cite{barends2014a, fowler2014, barends2014b, kelly2015, corcoles2015}. There is special interest in the study of the quantum-classical transition since decoherence is a technological issue in the realization of quantum computing, whose analysis leads to consideration of the interaction between system and environment.

Interest in OQS goes from studies of the influence of the environment on quantum properties in the quantum-classical transition \cite{caldeira1985,bedingham2013,hu1992,valdes2015,chou2008a,chou2008b}, to the study of the influence of the bath in uncertainty relations and the separation between quantum and thermal fluctuations \cite{abe1990, anderson1993, hu1993, hu1995, ponomarenko2001, isar2002}. Other important examples of OQS are photo-absorption of chromophores in a protein bath \cite{schroeder2006,tretiak2000,schroeder2007,ranger2010}, electron-phonon coupling in single-molecule transport \cite{burke2005,gebauer2005,gebauer2004,zheng2010,zheng2007,kurth2005} and exciton and energy transfer \cite{perdomo2010,rebentrost2010,rebentrost2009,abramavicius2004}.

Some of these studies have used the Shannon entropy from information theory as a tool to analyze the localization properties of the underlying distributions. There is also an increasing interest in the study of time-dependent information entropies \cite{dunkel2005, garbaczewski2005a, garbaczewski2005b, garbaczewski2006, haldar2013, aguiar2014}. The time-dependent behavior has been shown to be influenced by the initial data chosen for the dynamics \cite{garbaczewski2005a,aguiar2014}.

In this work, we analyze the influence of the bath on the localization properties of the distribution functions of the quantum system. We choose the harmonic oscillator and the Moshinsky atom \cite{mosha, moshb} as our model systems. The latter is a model of two indistinguishable particles subject to a harmonic oscillator potential coupled by another harmonic. We also study the interplay of the bath coupling and the interparticle interaction and their effects on the statistical correlation between particles. We examine the entropic uncertainty relations in both systems.

All these studies are carried out using the analytical solutions of the Lindblad master equation for the time evolution of the density operator. Particularly, we examine two limiting cases, the pure-dephasing without relaxation regime and the relaxation without pure-dephasing one.

\section{Information theory}

\subsection{Localization measures}

The Shannon entropy \cite{shannon,cover} is an uncertainty measure, hence a measure of the delocalization-localization of a distribution. This entropy can be defined for one-particle or two-particle distribution functions. It can also be defined for position and for momentum space probability distributions. If the distribution is time-dependent we can also define a time-dependent Shannon entropy. 
The one-particle Shannon entropy in position space is defined by

\begin{equation}
s_{x}(t) = -\int dx \ n(x,t) \ln{[n(x,t)]},
\label{sx}
\end{equation}
and measures the localization in $n(x,t)$, where $n(x,t)$, the position space density, is normalized to unity. The Shannon entropy increases when the underlying distribution delocalizes and decreases when it localizes. In this context, a larger entropy or uncertainty is associated with a loss of information while a smaller uncertainty is associated with a gain of information. 
In this manner, the Shannon entropies can be utilized to monitor changes in the localization-delocalization features of the underlying distributions.
It must be noted that the definition holds if the density comes from a one-particle system or if it is reduced by integration from a many-particle system with the appropriate normalization.

The corresponding uncertainty measure in momentum space is defined by,

\begin{equation}
s_{p}(t) = -\int dp \ n(p,t) \ln{[n(p,t)]},
\label{sp}
\end{equation}
with $n(p,t)$, the unity-normalized momentum space density.

As uncertainty measures, both Shannon entropies fulfill an entropic uncertainty principle \cite{bbm}, that for the one-dimensional case is,

\begin{equation}
s_{x} (t) + s_{p} (t) \geq 1 + \ln{\pi}.
\label{bound1}
\end{equation}
This entropic sum can be interpreted as an uncertainty measure of a separable phase-space distribution,
\begin{equation}
s_{t} (t) = -\int dx dp \ n(x,t) \ n(p,t)\ln{[n(x,t) \ n(p,t)]} = s_{x}(t) + s_{p}(t).
\label{st}
\end{equation}

Two-particle Shannon entropies, defined for position space as,

\begin{equation}
s_{x}^{2} (t) = \int dx_{1} dx_{2} \ n(x_{1},x_{2},t) \ln{[n(x_{1},x_{2},t)]}
\label{sx2}
\end{equation}
and for momentum space as

\begin{equation}
s_{p}^{2} (t) = \int dp_{1} dp_{2} \ n(p_{1},p_{2},t) \ln{[n(p_{1},p_{2},t)]},
\label{sp2}
\end{equation}
can be used to study the delocalization properties of the respective two-particle distribution functions.

There is an uncertainty principle at the two-particle level \cite{guevarajcp},

\begin{equation}
s_{x}^{2} (t) + s_{p}^{2} (t) \geq 2(1 + \ln{\pi})
\label{bound2}
\end{equation}
which can be interpreted as measuring the delocalization in the corresponding two-particle separable phase-space distribution,
\begin{equation}
s_{T} (t) = -\int dx dp \ n(x_{1},x_{2},t) \ n(p_{1},p_{2},t)\ln{[n(x_{1},x_{2},t) \ n(p_{1},p_{2},t)]} = s_{x}^{2}(t) + s_{p}^{2}(t).
\label{st2}
\end{equation}

\subsection{Correlation measures}

Mutual information \cite{shannon,cover} is defined in position and in momentum space \cite{sagar2005,sagar2006} as

\begin{equation}
I_{x}(t) = \int{ dx_{1} dx_{2} \ n(x_{1},x_{2},t) \ln \Bigg[{ \frac{n(x_{1},x_{2},t)}{n(x_{1},t) n(x_{2},t)} }} \Bigg] = 2 s_{x} (t) - s_{x}^{2} (t),
\label{mutinfx}
\end{equation}

\begin{equation}
I_{p}(t) = \int{ dp_{1} dp_{2} \ n(p_{1},p_{2},t) \ln \Bigg[{ \frac{n(p_{1},p_{2},t)}{n(p_{1},t) n(p_{2},t)} }} \Bigg] =2 s_{p} (t) - s_{p}^{2} (t).
\label{mutinfp}
\end{equation}
These quantities are measures of the correlation between positions or between momenta and are only zero when the two-variable distributions are separable. In all other cases they are greater than zero.

\section{Open quantum systems}

For the treatment of the OQS we start with the unitary evolution of the full density matrix (system plus reservoir),

\begin{equation}
\frac{d}{dt} \hat{\rho} (t) = -i {\bigg [} \hat{H}(t),\hat{\rho} (t) {\bigg ]}.
\end{equation}
The full Hamiltonian is given by
\begin{equation}
\hat{H}(t) = \hat{H}_{S} (t) + \hat{H}_{R} + \hat{V}
\end{equation}
where $\hat{H}_{S}(t)$ is the system Hamiltonian, $\hat{H}_{R}$ is the bath (reservoir) Hamiltonian, and $\hat{V}$ is the system-bath coupling, which is generally regarded to be weak. Defining the reduced density operator for the system alone by tracing over the reservoir degrees of freedom,

\begin{equation}
\hat{\rho}_{S} (t) = Tr_{R} \{ \hat{\rho}(t) \} ,
\end{equation}
one arrives at the formally exact quantum master equation,

\begin{equation}
\frac{d}{dt} \hat{\rho}_{S} (t) = -i {\bigg [} \hat{H}_{S} (t), \hat{\rho}_{S} (t) {\bigg ]} + \int_{t_{0}}^{t} d\tau \ \breve{\Xi} (t - \tau ) \hat{\rho}_{S} (\tau) + \Psi (t).
\label{mastereq}
\end{equation}
Here, $\breve{\Xi} (t - \tau )$ is the memory kernel and $\Psi (t)$ arises from initial correlations between the system and the environment. Although the equation is formally exact, in practice approximations are needed for $\breve{\Xi}$ and $\Psi$.

\subsection{Markov approximation and Lindblad master equation}

To make Eq. \ref{mastereq} easier to solve than the full equation, one often invokes the Markov approximation in which the memory kernel is local in time,

\begin{equation}
\int_{t_{0}}^{t} d\tau \ \breve{\Xi} (t - \tau ) \hat{\rho}_{S} (\tau) = \breve{D} \hat{\rho}_{S} (t).
\end{equation}
The Markov approximation is valid when the interaction between bath and system is weak \cite{tempel2011}.

The Lindblad form of the Markovian master equation,
\begin{equation}
\breve{D} \hat{\rho}_{S} (t) = \sum_{m,n} \bigg{\{} L_{mn} \hat{\rho}_{S}(t) L_{mn}^{\dagger} - \frac{1}{2} L_{mn}^{\dagger} L_{mn} \hat{\rho}_{S}(t) - \frac{1}{2} \hat{\rho}_{S}(t) L_{mn}^{\dagger} L_{mn} \bigg{\}} ,
\end{equation}
is constructed to guarantee complete positivity of the density matrix \cite{oqs1}.
Two limiting cases of the above Lindblad equation: pure-dephasing without relaxation and relaxation without pure-dephasing will be discussed in the next section.

\section{Models studied}

In this section, we will discuss the results of our analysis to two ubiquitous models whose physics is well understood and that can provide the physical intuition to tackle more complex systems in the future. We begin by describing the harmonic oscillator and continue by discussing the Moshinsky atom.

\subsection{Harmonic oscillator}

For the 1D harmonic oscillator (HO) Hamiltonian (atomic units ($m= \hbar = 1$) are used throughout the paper),

\begin{equation}
\hat{H}_{S} = -\frac{1}{2} \frac{d^{2}}{dx^{2}} + \frac{1}{2}\omega^{2} x^{2},
\end{equation}
the Lindblad master equation

\begin{equation}
\frac{d \hat{\rho}_{S} (t)}{dt} = -i [\hat{H}_{S} , \hat{\rho}_{S}(t)] + \sum_{m,n} \bigg{\{} L_{mn} \hat{\rho}_{S}(t) L_{mn}^{\dagger} - \frac{1}{2} L_{mn}^{\dagger} L_{mn} \hat{\rho}_{S}(t) - \frac{1}{2} \hat{\rho}_{S}(t) L_{mn}^{\dagger} L_{mn} \bigg{\}}
\label{lindblad_eq}
\end{equation}
can be solved analytically by obtaining the coefficients of the density operator

\begin{equation}
\hat{\rho}_{S}(t) = \sum_{n,m = 0} ^{1} \rho_{n,m}(t) \ | n \rangle \langle m | .
\label{dens_op}
\end{equation}

Upper and lower limits in the summation are a result of considering only the first two eigenstates of HO. This approximation is valid in the low temperature regime for the relaxation case and because the initial condition contains only the first two states for the pure-dephasing case. We assume the initial state to be formed by the first two eigenstates,

\begin{equation}
| \psi (0) \rangle = \frac{1}{\sqrt{2}} (| 0 \rangle + | 1 \rangle )
\end{equation}
which is equivalent to consider $\rho_{0,0} (0) = \rho_{0,1} (0) = \rho_{1,0}(0) = \rho_{1,1}(0) = \frac{1}{2}$. This model was discussed in the context of TDDFT for open quantum systems \cite{yuen2010, tempel2012, yuen2013, yuen2009} in Ref. \cite{tempel2011}.

At this point, we can express the density operator in its position or momentum representation by taking

\begin{eqnarray}
n(x,t) = \langle x | \hat{\rho}_{S}(t) | x \rangle \\
n(p,t) = \langle p | \hat{\rho}_{S}(t) | p \rangle.
\end{eqnarray}

\subsubsection{Pure-dephasing without relaxation}

Pure-dephasing describes a situation in which the system-bath collisions are elastic so that the bath decoheres the system without exchanging energy. In this regime, the Lindblad operators take the form,

\begin{equation}
L_{mn} = \delta_{mn} \sqrt{\frac{\gamma_{m}}{2}} | m \rangle \langle m |
\end{equation}
where $\delta_{mn}$ is the Dirac delta function.

Using these operators and equation ({\ref{dens_op}) yields populations,
\begin{equation}
\rho_{00}(t) = \rho_{00}(0) = \frac{1}{2}
\end{equation}
and
\begin{equation}
\rho_{11}(t) = \rho_{11}(0) = \frac{1}{2}.
\end{equation}
This means that there is no relaxation in the system and the energy is not affected by the bath. For the coherences, we have
\begin{equation}
\rho_{01}(t) = \rho_{01}(0) e^{-i(E_{0} - E_{1}) t} e^{-\frac{1}{2}(\frac{\gamma_{0} + \gamma_{1}}{2}) t } = \frac{1}{2} e^{i \omega t} e^{-\frac{1}{2}(\frac{\gamma_{0} + \gamma_{1}}{2}) t }
\end{equation}
and
\begin{equation}
\rho_{10}(t) = \rho_{10}(0) e^{-i(E_{1} - E_{0}) t} e^{-\frac{1}{2}(\frac{\gamma_{0} + \gamma_{1}}{2}) t } = \frac{1}{2} e^{-i \omega t} e^{-\frac{1}{2}(\frac{\gamma_{0} + \gamma_{1}}{2}) t }.
\end{equation}
They oscillate and decay exponentially due to the presence of the bath. For the calculations we take $\gamma_{0} =\gamma_{1}$ and $\gamma_{0} \equiv \gamma$, to simplify the notation.

\subsubsection{Relaxation without pure-dephasing}

In this case the Lindblad operators are strictly non-diagonal

\begin{eqnarray}
\nonumber
L_{mn} &= \sqrt{\gamma_{mn}} | m \rangle \langle n | & \qquad \text{if} \qquad m \neq n \\
L_{mn} & = 0 & \qquad \text{if} \qquad m = n
\end{eqnarray}

The populations evolve according to

\begin{equation}
\frac{d}{dt} \rho_{nn} (t) = \sum_{m} \gamma_{nm} \ \rho_{mm} (t) - \rho_{nn}(t) \sum_{m} \gamma_{mn}.
\end{equation}
The first term is the rate at which population leaves from $\rho_{nn}$, and the second term is the rate at which population is transferred to $\rho_{nn}$ from the other states. Considering the first two levels, we get the populations from solving the coupled differential equations subject to the above mentioned initial conditions.
Thus

\begin{eqnarray}
\frac{d}{dt} \rho_{00} (t) = \gamma_{01} \ \rho_{11}(t) - \gamma_{10} \ \rho_{00}(t) \\
\frac{d}{dt} \rho_{11} (t) = \gamma_{10} \ \rho_{00}(t) - \gamma_{01} \ \rho_{11}(t).
\end{eqnarray}
The coherences evolve as
\begin{equation}
\rho_{nm}(t) = \rho_{nm}(0) e^{-i(E_{n} - E_{m}) t} e^{-\frac{1}{2}(\sum_{l} \gamma_{ln} + \sum_{l} \gamma_{lm}) t}
\end{equation}
and leads to

\begin{eqnarray}
\rho_{01}(t) = \frac{1}{2} e^{i\omega t} e^{-\frac{1}{2}(\gamma_{10} + \gamma_{20} + \gamma_{01} + \gamma_{21}) t} \\
\rho_{10}(t) = \frac{1}{2} e^{-i \omega t} e^{-\frac{1}{2}(\gamma_{10} + \gamma_{20} + \gamma_{01} + \gamma_{21}) t},
\end{eqnarray}
which means that the bath decoheres the system even in absence of pure-dephasing. In fact, the relaxation regime is closer to a real physical situation 
since one expects that relaxation and decoherence occur simultaneously when the system interacts with the bath.

We can approximate all the $\gamma$'s in terms of $\gamma_{01}$ which governs the rate of the relaxation from the state $| 1 \rangle$ to the state $| 0 \rangle$ (which is the largest possible relaxation rate under these assumptions) by means of the relations (with $\omega =1$),
\begin{eqnarray}
\nonumber
\gamma_{10} & = \gamma_{01} e^{-(E_{1} - E_{0})} = \gamma_{01} e^{-\omega} & = \frac{\gamma_{01} }{e} , \\
\nonumber
\gamma_{20} & = \gamma_{02} e^{-(E_{2} - E_{0})} = \gamma_{02} e^{-2\omega} & \approx \frac{\gamma_{01} }{e^{2}} , \\
\gamma_{21} & \approx \gamma_{10} & = \frac{\gamma_{01} }{e}.
\end{eqnarray}
Thus, it is sufficient to specify the value of $\gamma_{01}$ and we use $\gamma_{01} \equiv \gamma$ in order to simplify notation.

\section{Moshinsky atom}

We now proceed to study the 1D Moshinsky atom repulsive Hamiltonian \cite{mosha,moshb},

\begin{equation}
\hat{H}_{S} = -\frac{1}{2} \frac{d^{2}}{dx_{1}^{2}} - \frac{1}{2} \frac{d^{2}}{dx_{2}^{2}} + \frac{1}{2}\omega^{2} x_{1}^{2} + \frac{1}{2}\omega^{2} x_{2}^{2} - \frac{1}{2} \lambda (x_{1} - x_{2})^{2} ,
\end{equation}
the Lindblad master equation,

\begin{eqnarray}
\nonumber
\frac{d \hat{\rho}_{S} (t)}{dt} = & - i [\hat{H}_{S} , \hat{\rho}_{S}(t)] + \sum_{mn,m^{\prime}n^{\prime}} \bigg{\{} L_{mn,m^{\prime}n^{\prime}} \hat{\rho}_{S}(t) L_{mn,m^{\prime}n^{\prime}}^{\dagger} \\
& - \frac{1}{2} L_{mn,m^{\prime}n^{\prime}}^{\dagger} L_{mn,m^{\prime}n^{\prime}} \hat{\rho}_{S}(t) - \frac{1}{2} \hat{\rho}_{S}(t) L_{mn,m^{\prime}n^{\prime}}^{\dagger} L_{mn,m^{\prime}n^{\prime}} \bigg{\}}
\label{lindblad_eq_mosh}
\end{eqnarray}
can be solved analytically by obtaining the coefficients of the density operator

\begin{equation}
\hat{\rho}_{S}(t) = \sum_{m,n,m^{\prime},n^{\prime} = 0} ^{1} \rho_{mn,m^{\prime}n^{\prime}}(t) \ | mn \rangle \langle m^{\prime}n^{\prime} |.
\label{dens_op_mosh}
\end{equation}
$| mn \rangle = | m \rangle | n \rangle$ are the eigenstates of the Moshinsky atom. In this notation, $m$ is associated with the center-of-mass coordinates and $n$ with the relative ones. The relative coordinates govern the symmetry of a particular solution. For $n$ even, the wave functions are symmetric whereas for $n$ odd they will be antisymmetric. Upper and lower limits in the summation come from the fact that we consider just the first two eigenstates. Accordingly, we assume the initial state,

\begin{equation}
| \psi (0) \rangle = \frac{1}{\sqrt{2}} (| 00 \rangle + | 10 \rangle ).
\end{equation}
Both eigenstates have the same symmetry. They are symmetric under the interchange of the original coordinates. This choice is equivalent to $\rho_{00,00} (0) = \rho_{00,10} (0) = \rho_{10,00}(0) = \rho_{10,10}(0) = \frac{1}{2}$.

For considering the pure-dephasing without relaxation and the relaxation without pure-dephasing regimes we proceed in complete analogy with the HO model presented above.

\section{Harmonic oscillator coupled with the bath}

The Shannon entropies, Eqs. (\ref{sx}) and (\ref{sp}), of the HO coupled to a bath are shown in Fig. (\ref{fig1}) for both regimes: the pure-dephasing without relaxation and the relaxation without pure-dephasing, with the coupling parameters $\gamma =0.15$, $\gamma =0.3$ and $\gamma=0.5$.

The interpretation of the curves in Fig. (\ref{fig1}) is that the delocalization-localization behavior is damped with time and this damping depends on the strength of the bath coupling constant (on comparing the curves for different values of $\gamma$). 

For the pure-dephasing without relaxation regime, the dynamics in position space starts with a highly localized state at $t=0$ (smaller Shannon entropy), where we have more information about the quantum system. As the system evolves in time, the entropy increases, i.e. the particle becomes more delocalized in successive periods of the evolution and asymptotically approaches the thermal equilibrium value (the mixed state at $t = \infty$), where the coherences are zero. The concept here is that the state delocalizes in x-space and loses information as the coherences diminish. This behavior can be observed from the densities in Fig. (\ref{fig2}). The density at $t=0$ is highly localized. It delocalizes at $t=\pi/2$ and localizes again at $t=\pi$ and $t=2\pi$ (but delocalized as compared with $t=0$).

In momentum space, the difference is that at $t=0$ we start with a highly delocalized state (less information) whose entropy is the same as that of the thermal equilibrium. However note that in the evolution of the system we periodically gain and lose information.

The coherences are zero at half intervals of $\pi$. These points are characterized by a maximum entropy in position space (delocalization or loss of information). 
At these points, there is no difference in the Shannon entropies corresponding to different bath couplings as the densities are the same for these points (see Fig. (\ref{fig2})). That is, the delocalization (loss of information) for different intensities of coupling with the bath is the same. For other points in the period, the loss of information provoked by a larger coupling with the bath is greater since its Shannon entropy is larger.

In momentum space, the zero coherences at half intervals of $\pi$ are characterized by localization or gain of information. The larger coupling to the bath is characterized by loss of information since the Shannon entropy is larger. Thus, the off-diagonal or quantum interference terms must serve to localize the density in position space and delocalize it in momentum space. 

In the case of the relaxation without pure-dephasing regime, we start in position space with a highly localized state and we lose information with the system's evolution, this behavior of the density is apparent in Fig. (\ref{fig2}). In momentum space, we start with a highly delocalized state, but the difference with pure-dephasing is that the value at $t=0$ is not the thermal equilibrium value. The Shannon entropy at thermal equilibrium is smaller than at $t=0$, thus we have gained information as compared to this initial point.

At half intervals of $\pi$ in the relaxation regime, the larger interaction with the bath leads to smaller loss of information in position space, but larger in momentum space. Note also that this behavior is reversed at integer values of $\pi$, $t= n\pi$. A larger interaction leads to larger loss of information in x-space and a smaller loss of information in p-space. For the position space we can interpret this behavior by observing that of the densities in Fig. (\ref{fig2}).

Remarkably, in the pure-dephasing regime, the entropic sum $s_{t}$ is sensitive to the strength of the interaction with the bath, $s_{t}^{\gamma =0.5} > s_{t}^{\gamma =0.3} > s_{t}^{\gamma =0.15}$. The entropy sum has been shown to be sensitive to the effects of interparticle repulsions \cite{guevara,lagunapra} in a closed quantum system. 
These results illustrate that it is also sensitive to the effects of the coupling with the environment. A larger coupling to the bath provokes a larger entropic sum. 
These results are also valid in the relaxation regime for $t < \pi$, where additionally the entropy sums reach a maximum value and then decrease. This means that the separable phase-space density first delocalizes (loss of information) for smaller $t$ and then localizes (information gain) for larger values of $t$. For greater values of $t$, there are crossovers in the curves.
The physical interpretation of these crossovers is that the phase-space density is more delocalized when the coupling to the bath is stronger, at smaller values of time. This behavior is inverted when the system approaches the equilibrium state. It is the weakest coupled which is most delocalized.  Hence, at equilibrium, the information loss in phase-space is greater when the system-bath coupling is smaller.

\section{Moshinsky atom coupled with a bath}

\subsection{The non-interacting case}

Setting the interparticle potential equal to zero ($\lambda=0$), the Moshinsky atom becomes a system of two noninteracting oscillators, whose wave function is symmetric. In this way, we study the effects of the bath on the system and eliminate any effects due to the interparticle potential.

In Fig. (\ref{fig3}) we plot the entropies in both regimes and for several values of $\gamma$. We observe that the behavior is similar to that of the HO case. In the pure-dephasing regime in x-space, a larger coupling to the bath leads to a larger loss of information in general. In the relaxation regime, there is a more complicated behavior which alternates between a larger loss of information with increasing bath coupling strength and the inverse behavior (smaller loss of information with increasing coupling strength). This depends on the particular value of $t$. 

The position space Shannon entropy increases from $t=0$ to $t=\frac{\pi}{2}$. At this latter point the coherences are zero. In the pure-dephasing regime, this point is characterized by a loss of information which does not depend on the coupling strength. In momentum space, this point is characterized by a gain of information (smaller entropy-more localized distribution) which does depend on the coupling strength. A smaller coupling strength induces a larger gain in information. 
In the relaxation regime the behavior is similar but in position space the loss of information now depends on the coupling strength. A larger coupling results in smaller gain of information in momentum space.

The entropic sum which measures the uncertainty in a separable phase-space distribution increases with $t$ and is sensitive to the coupling strength of the bath. A larger coupling strength yields a larger entropy in the pure-dephasing regime. In the relaxation regime, this result holds for $t < \pi$. For greater $t$, there are crossovers, and the entropy sums display maxima with the same physical interpretation as in the HO case. 

\subsection{The effect of the interparticle potential}

\subsubsection{One-variable Shannon entropies}

In this section we examine the effect of the interparticle potential in the presence of the bath, on the localization of the distribution functions and in the correlation measures. We start with the analysis of the reduced one-variable Shannon entropies.

In Fig. (\ref{fig4}) we plot the entropies for both regimes. A larger interparticle potential provokes a larger $s_{x}$, hence larger loss of information. In momentum space the behavior is opposite. Larger $\lambda$ yields a smaller $s_{p}$ (localization). This occurs in both regimes. These effects are the same as those observed for the time-independent closed system \cite{lagunapra}.

The plots of $s_t$ show that the entropic sum tends toward asymptotic equilibrium values.
We also observe intersections between the plots in the vicinity of $t = \pi /2$. Also interesting is that the plots show more structure for larger values of the interparticle interaction, $\lambda$. This structure is more pronounced in the relaxation without pure-dephasing regime where we observe maxima in all plots. These maxima occur in the first period of the evolution $t \in [0,2 \pi]$. Beyond this period all curves are characterized by a monotonically decreasing behavior.
Note also that all points are above the bound for $s_t$ in Eq. (\ref{bound1}).

The entropic sum as a function of the interparticle potential is plotted in Fig. (\ref{fig5}) for both regimes and for different values of time. The same increasing trend is displayed in most of the plots. However there are minima for $t= \pi /2$ in both regimes. This minimum is more pronounced in the pure-dephasing regime. For $t= \pi /2$, the coherences, are zero in both regimes for the first time in the evolution. At this point, the momentum space distribution localizes more strongly than the delocalization in the position space distribution, which is an effect of the potential. The fact that these minima do not occur again for other half values of $\pi$ can be interpreted as an effect due to the interaction with the bath, which decoheres the system with time.

\subsubsection{Two-variable Shannon entropies}

We now turn our attention to the two-variable non-reduced distributions and their entropies, in the presence of the bath. The effect of the potential is the same as discussed above for the reduced entropies. As observed from Fig. (\ref{fig6}), a stronger interparticle potential induces a larger delocalization in the position space distribution (information loss) and a larger localization in the momentum space one (information gain).

The most striking difference is the behavior of the entropic sum in the different regimes. In the pure-dephasing without relaxation regime, $s_T$ does not depend on the interparticle potential while it does in the relaxation without pure-dephasing regime. All points are above the bound for $s_T$ in Eq. (\ref{bound2}).

In Fig. (\ref{fig7}), the interparticle potential dependence of the entropic sum is plotted. $s_{T}(\lambda)$ is a constant for the pure-dephasing regime. This is also a characteristic of the (time independent) closed quantum system \cite{lagunapra}.

On the other hand, the dependency of the entropic sum on the interparticle potential in the relaxation without pure-dephasing regime comes from the population changes with the interaction with the bath. Thus, the increasing behavior of the sum with the potential is the result of delocalization in position space which is not compensated with an equal localization in momentum space. Hence the effect of the bath is greater in position space than in momentum space. 
This is different from the pure-dephasing regime where delocalization in position space is perfectly compensated by localization in momentum space. In this case, the effect of the bath is the same in both spaces.

\subsubsection{Mutual information}

In this section, we discuss the combined effect of the bath and the interparticle potential on the statistical correlation between particles.

In Fig. (\ref{fig8}) we present the correlation between particles measured in position and in momentum space, and their sum (the total correlation) for $\gamma=0.15$. The behavior is similar for other values of the bath coupling.

In position space, the statistical correlation between particles is smaller with larger interparticle potential at larger values of $t$. The most statistically correlated case is in fact the non-interacting system. This behavior, contrary to intuition, is a decoherence effect due to the bath. 
Details of the smaller $t$ part of the plot are shown in Fig. (\ref{fig9}). The dynamics starts with the expected behavior at $t=0$. That is, the most correlated case is the one with the larger interparticle potential. However, this correlation rapidly decays and the intersections in the plots are apparent. Thus, the effect of the bath makes the more repulsively correlated system less statistically correlated. That is, the interaction with the bath decouples the positions of the particles as the system evolves towards equilibrium. This behavior also holds for different values of the system-bath coupling strength. Thus, for small $t$, the statistical correlation is governed by the interparticle potential whereas for greater $t$, it is governed by the interaction with the bath. This behavior is present in both regimes. The difference between the two is that
the asymptotic values of the correlation at large $t$ is smaller for the relaxation regime as compared to the pure-dephasing one for a particular value of $\lambda$.
This effect is due to the changing populations in the relaxation regime.

On the other hand, the momenta of the particles are more statistically correlated for larger values of the interparticle potential as expected. The structure of the plots is essentially the same for the different values of $\lambda$. 

The sum of the correlation between positions and between momenta, $I_t$, can be interpreted as the total correlation between the particles in the system. As shown in the plots, the total correlation is larger for the greater values of the interparticle potential for the most part of the evolution. There are, however, intersections between the curves that occur at small $t$. This behavior illustrates the interplay between the interparticle potential and the interaction with the bath, in governing the total correlation between particles. Details of this interplay can be observed in Fig. (\ref{fig9}).

Another question to be asked is if the strength of the coupling to the bath affects the statistical correlation between particles in the same manner
as the strength of the interparticle repulsive potential. To analyze this, we present plots of the correlation for the non-interacting system ($\lambda =0$) at different values of the coupling strength $\gamma$ in Figs. (\ref{fig10}) and (\ref{fig11}). At $t=0$, the correlation is the same in all instances and stems from the indistinguishability of the particles and the quantum superposition.

In position space, at very small $t$, the effect of a stronger interaction with the bath is a smaller correlation in the pure-dephasing regime. This behavior is similar to that previously observed for the interparticle potential in Figs. (\ref{fig8}) and (\ref{fig9}) at larger $t$. The difference here is that for $t > \frac{\pi}{4}$, the curves intersect and the more strongly coupled to the bath is now the more correlated case. In the relaxation regime, a stronger coupling to the bath yields a weaker correlation for all $t$. This behavior is the same as that between the correlation and interparticle potential in Figs. (\ref{fig8}) and (\ref{fig9}) at larger $t$. 

In the pure-dephasing regime in momentum space, a stronger coupling to the bath results in a stronger correlation between momenta. This relation is the same as that observed between the interparticle potential and the correlation in Fig. (\ref{fig8}). In the relaxation regime, a stronger coupling to the bath yields a weaker correlation and is the same trend as that observed in position space. This trend is opposite to that in Fig. (\ref{fig8}) where a larger interparticle potential provokes a stronger correlation.

The total correlation in the pure-dephasing regime increases with a stronger coupling to the bath. This behavior is in general consistent with that of Fig. (\ref{fig9}), except for values close to $t=0$. On the other hand, this behavior is inverted in the relaxation regime which is expected since both its components display the same behavior.

In Fig. (\ref{fig12}), we systematically compare the correlations between positions with the correlations between momenta as measured by the mutual information. For the non-interacting case, the values of the position and the momentum correlation are closer than in the interacting cases ($\lambda =0.3, 0.5$). Thus, the effect of the interparticle potential is to separate $I_{x}$ from $I_{p}$. When the interparticle potential is turned on ($\lambda \neq 0$), the statistical correlation between the momenta is larger than the correlation between positions. These trends also hold for other values of the bath coupling and are not presented for brevity.

\section{Conclusions}

The effects of the interaction between a two-particle quantum system and its environment is studied by the use of the Moshinsky atom model and the Lindblad operator
master equation formalism. The purpose of this work is to gauge the response of the system to the coupling with the environment, and the interplay with the interparticle repulsive potential. We use Shannon entropies to examine how the environmental coupling and interparticle potential affect the localization-delocalization features of the one (reduced) and two-particle distribution functions in position, momentum and separable phase-space. Furthermore, mutual information is used to study the statistical correlation between particles and how this correlation is affected by the environment and the repulsive potential. Bath coupling in position space leads to a delocalization of the one and two-particle distribution functions, which can be interpreted as a loss of information. On the other hand, bath coupling in momentum space leads to a localization of the underlying distributions, which can be interpreted as a gain of information. The (reduced) one-particle entropy sum measures the delocalization of a separable phase-space distribution and forms the basis of the entropic uncertainty relation. Results show that in the presence of the bath it is sensitive to and increases with the strength of the interparticle potential. The only point where this is not valid is at $t=\pi/2$ where the off-diagonal elements of the density matrix are zero. At this point, the entropy sum is shown to transit through a minimum as the potential is increased. The entropy sum is also shown to be sensitive to the strength of the bath coupling. The two-particle entropy sum is constant and does not depend on the strength of the interparticle potential ($\lambda$) in the pure-dephasing regime. On the other hand, it increases with $\lambda$ in the relaxation regime. The interpretation of this behavior is a loss of information in a separable phase-space distribution.

The statistical correlation between the particles positions and between their momenta, in the presence of the bath, display markedly different behaviors. At $t=0$, the magnitude of the correlation in position space depends on the strength of the interparticle potential. That is, the positions of the particles are more statistically correlated in the system with the strongest interparticle potential. However, in the presence of the bath, it is the system with weakest interparticle potential that has the largest statistical correlation. Thus the effect of the bath is to randomize the behavior of the particles which yields a smaller correlation between their positions. Hence we conclude that the coupling to the bath governs the statistical correlation between particles for larger values of $t$, whereas the interparticle potential is responsible for this correlation at small $t$. The point taken from this analysis of the position space mutual information is that the bath decouples the particles. It would be interesting to investigate if this effect is maintained for other potentials such as the Coulombic one for the case of realistic electronic systems.

On the contrary, particles with a stronger interparticle potential are more correlated with regard to their momenta, even in the presence of the bath. Based on this, the observation is that the effects of the bath on the statistical correlation between the particles is more pronounced in position space as compared to momentum space. We also find that the presence of a repulsive interparticle potential in a system that is coupled to a bath yields that the magnitude of the statistical correlation between the particles momenta is greater than that between their positions. The difference between these magnitudes increases with the strength of the potential.

The two-particle model studied in this work can be generalized to $N$ particles, and is still analytically solvable. Thus one can study many-body effects on the statistical correlation between two particles and its interplay with the bath coupling. Employing suitable definitions of higher-order mutual information, one could examine other collective effects present in this open quantum system and test if the observed trends about information gain/loss and statistical correlations persist for $N$ particle systems.

\section*{Acknowledgements}

H. G. L. thanks DGAPA-UNAM for a postdoctoral fellowship. H. G. L. also thanks CONACyT for a Beca Mixta and the Aspuru-Guzik group for their hospitality. A. A.-G. and D. G. T. acknowledge support from the National Science Foundation under award CHE-1464862 \& CISE/ACI CIF21. The authors thank CONACyT for travel support enabling D. G. T.'s visit to Mexico City.

\newpage

\section{Figures}

\begin{figure}[H]
\begin{center}
\includegraphics{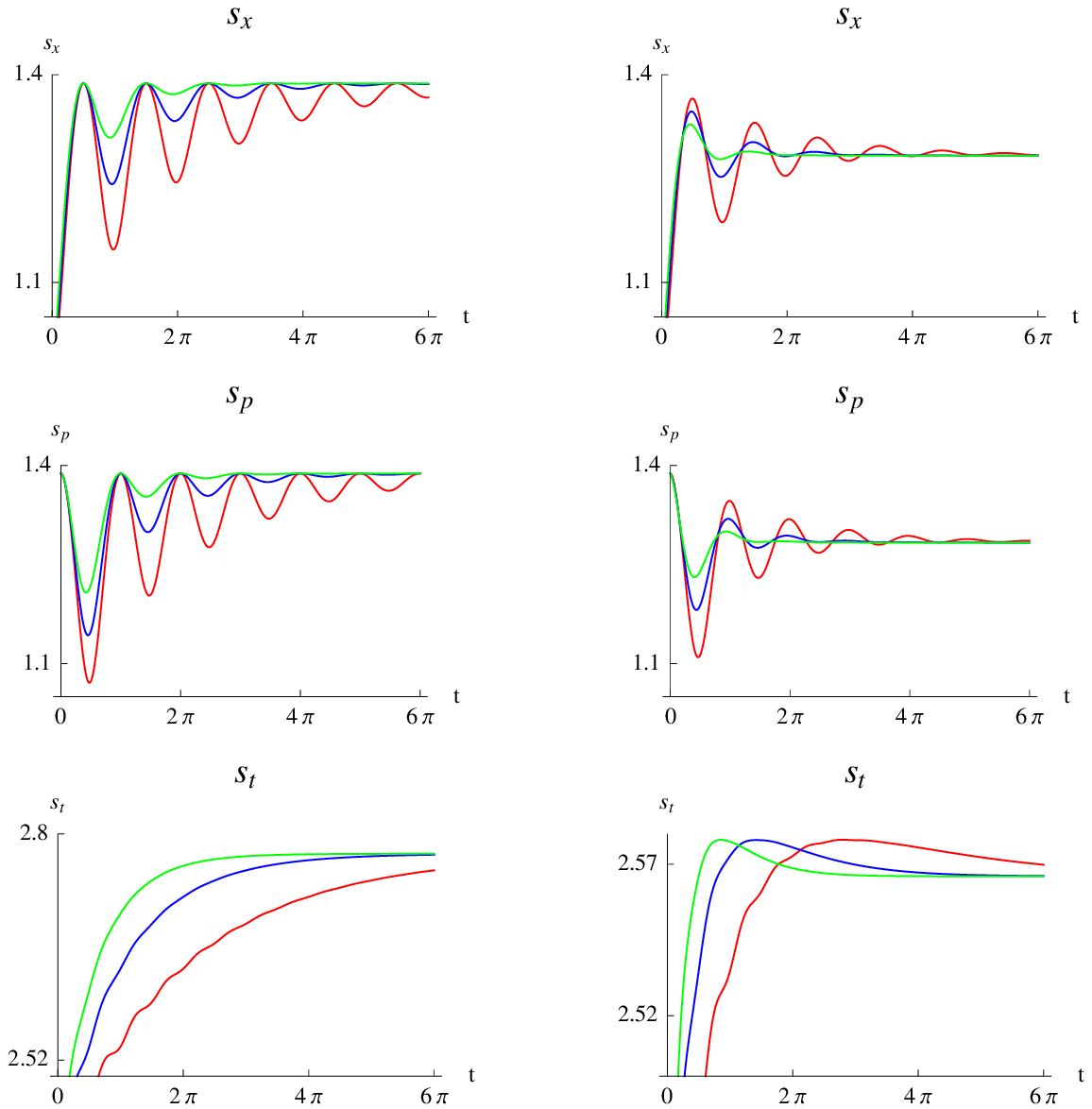}
\caption{Position space entropy as a function of time, $s_{x}(t)$, Eq. (\ref{sx}) (top), momentum space entropy as a function of time, $s_{p}(t)$, Eq. (\ref{sp}) (middle) and total entropy as a function of time, $s_{t} (t)$ (bottom) of the harmonic oscillator, for the case of pure-dephasing without relaxation (left) and relaxation without pure-dephasing (right). $\omega=1$ and $\gamma =0.15$ (red), $\gamma =0.3$ (blue), $\gamma =0.5$ (green).}
\label{fig1}
\end{center}
\end{figure}

\begin{figure}[H]
\begin{center}
\includegraphics{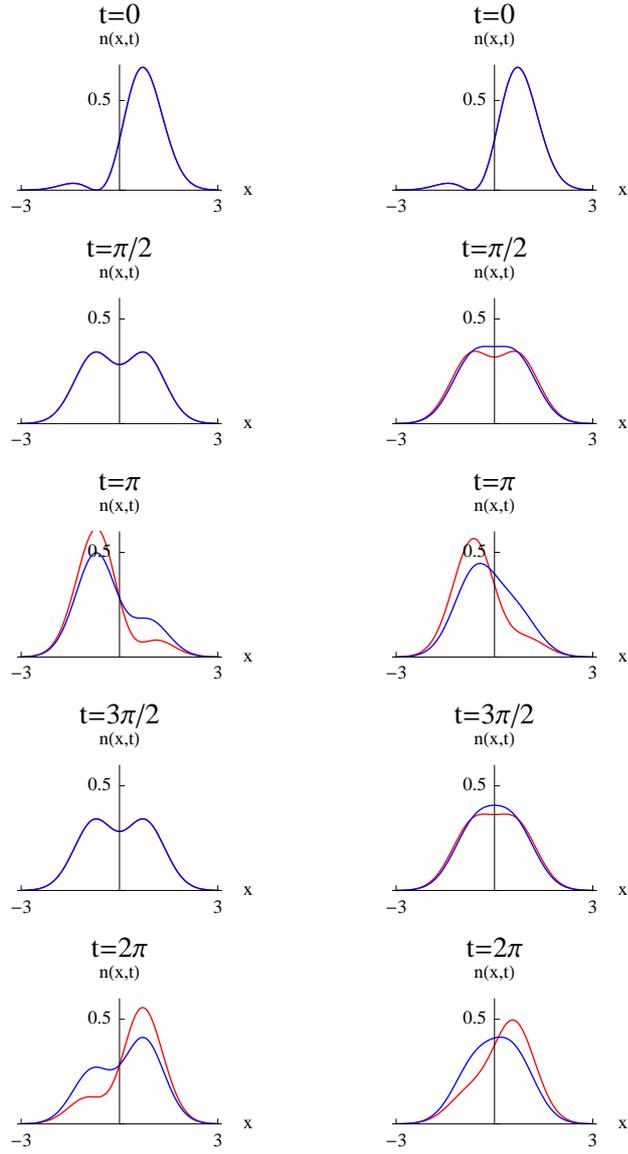}
\caption{Position space density, $n(x,t)$, of the harmonic oscillator for the case of pure-dephasing without relaxation (left) and relaxation without pure-dephasing (right). $\omega=1$ and $\gamma =0.15$ (red), $\gamma =0.5$ (blue).}
\label{fig2}
\end{center}
\end{figure}

\begin{figure}[H]
\begin{center}
\includegraphics{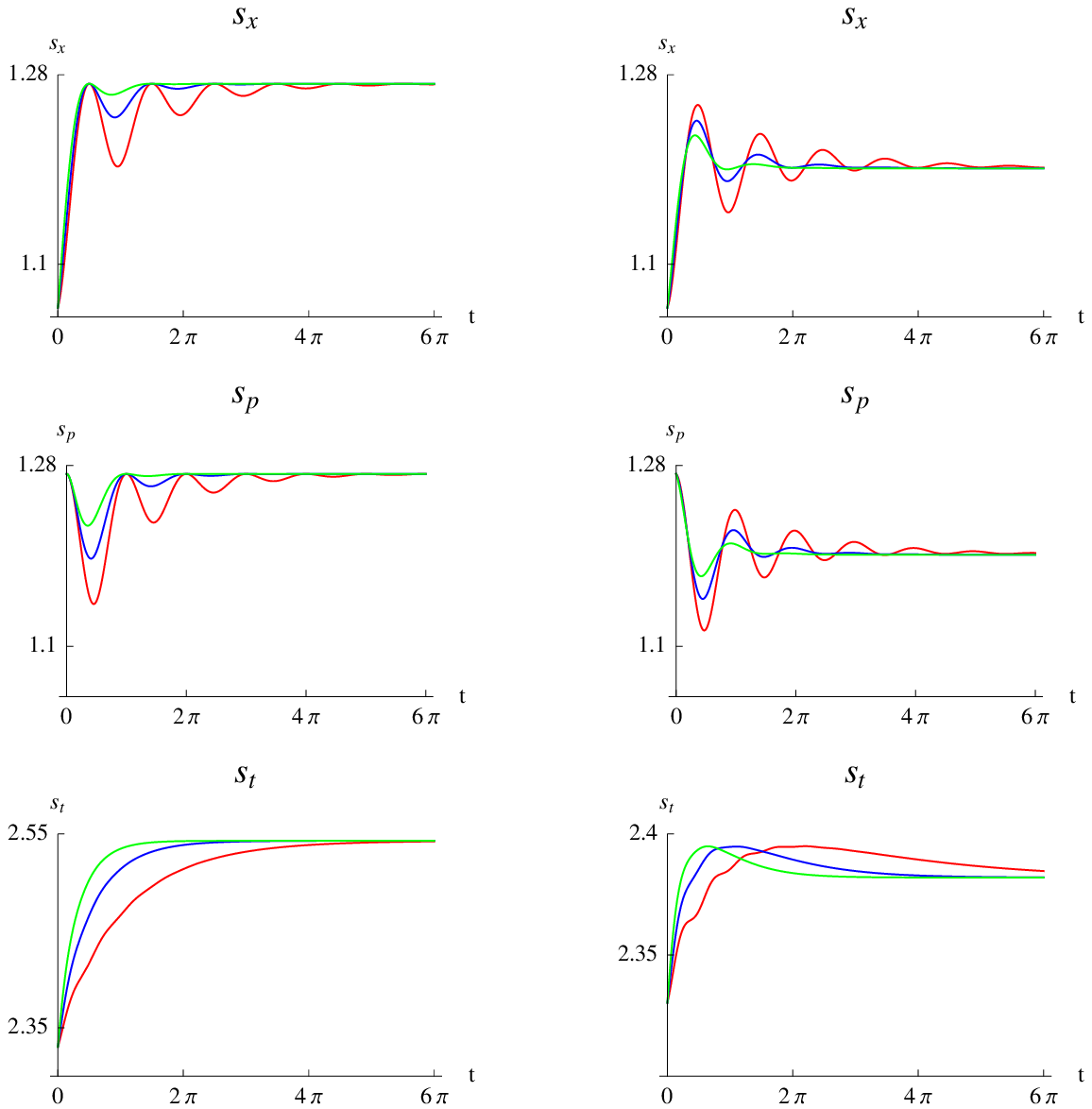}
\caption{Position space entropy as a function of time, $s_{x}(t)$, Eq. (\ref{sx}) (top), momentum space entropy as a function of time, $s_{p}(t)$, Eq. (\ref{sp}) (middle) and total entropy as a function of time, $s_{t} (t)$ (bottom), for the case of pure-dephasing without relaxation (left) and relaxation without pure-dephasing (right) in the non-interacting Moshinsky atom ($\lambda = 0$). $\omega=1$ and $\gamma =0.15$ (red), $\gamma =0.3$ (blue), $\gamma =0.5$ (green).}
\label{fig3}
\end{center}
\end{figure}

\begin{figure}[H]
\begin{center}
\includegraphics{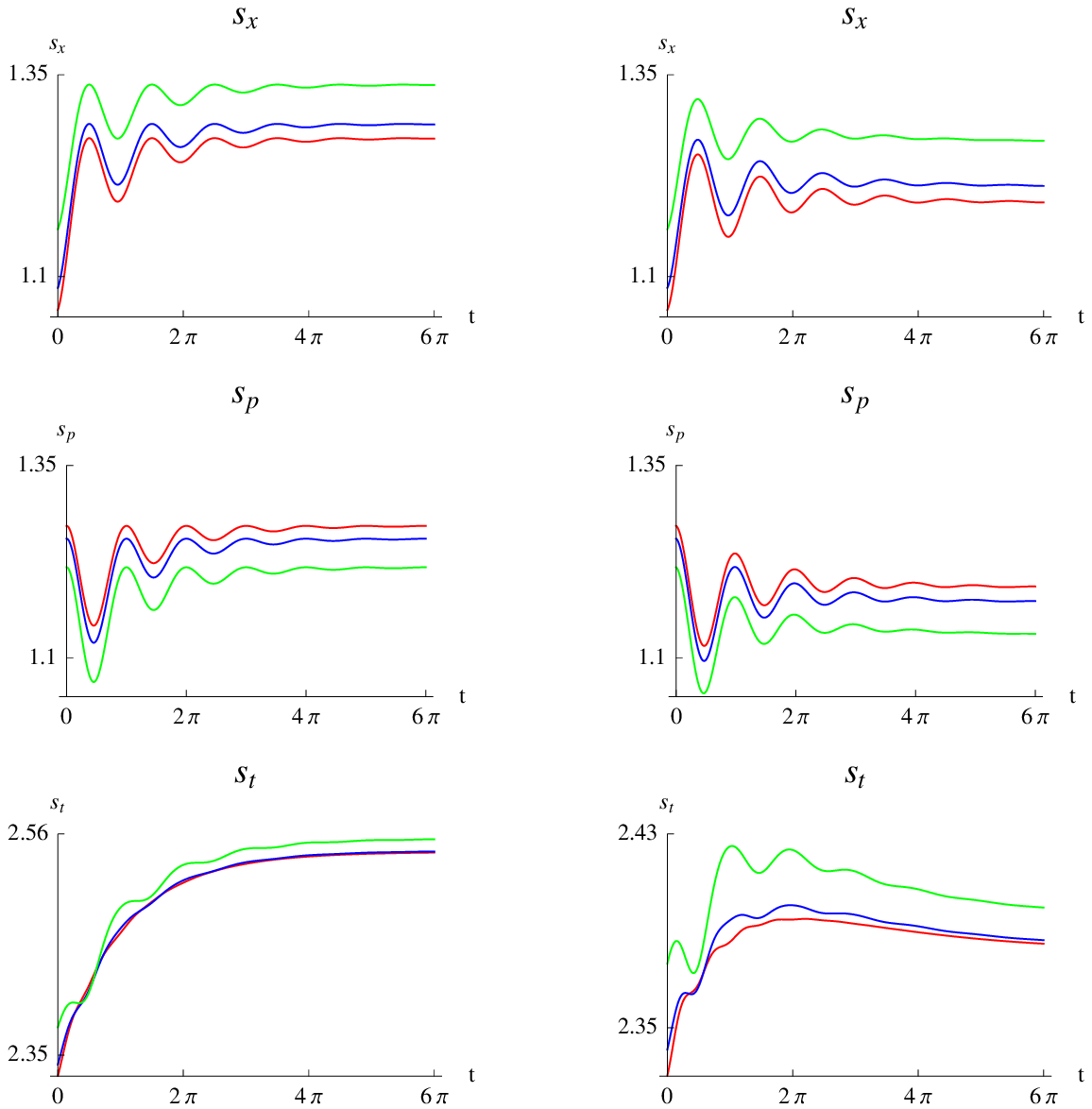}
\caption{Position space entropy as a function of time, $s_{x}(t)$, Eq. (\ref{sx}) (top), momentum space entropy as a function of time, $s_{p}(t)$, Eq. (\ref{sp}) (middle) and total entropy as a function of time, $s_{t} (t)$ (bottom), of the Moshinsky atom, for the case of pure-dephasing without relaxation (left) and relaxation without pure-dephasing (right). $\omega=1$ and $\gamma =0.15$. $\lambda = 0$ (red), $\lambda = 0.3$ (blue), $\lambda = 0.5$ (green).}
\label{fig4}
\end{center}
\end{figure}

\begin{figure}[H]
\begin{center}
\includegraphics{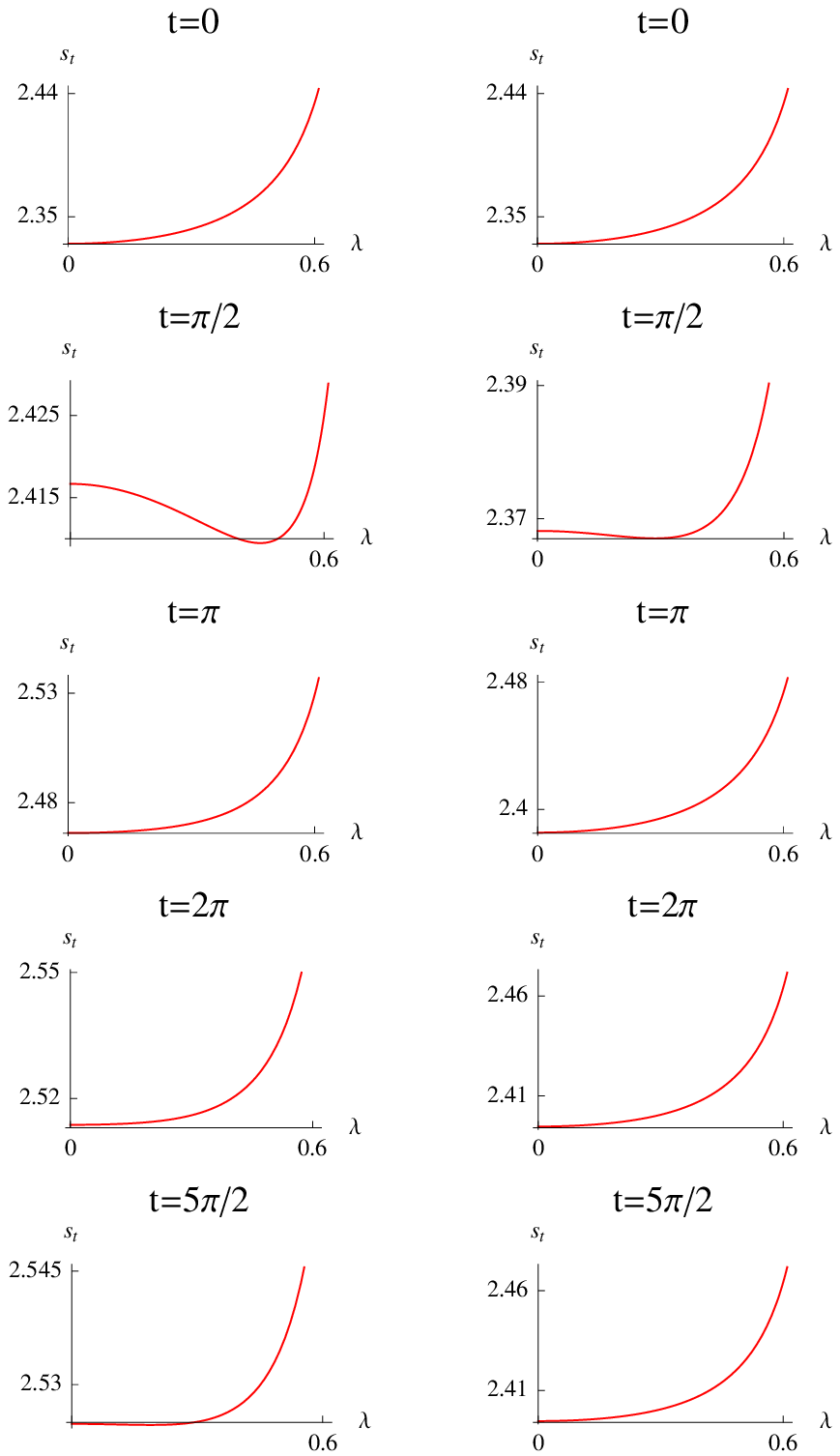}
\caption{Total entropy as a function of the interparticle potential, $s_{t} (\lambda)$, of the Moshinsky atom, for the case of pure-dephasing without relaxation (left) and relaxation without pure-dephasing (right). $\omega=1$ and $\gamma =0.15$, for different values of time, $t$.}
\label{fig5}
\end{center}
\end{figure}

\begin{figure}[H]
\begin{center}
\includegraphics{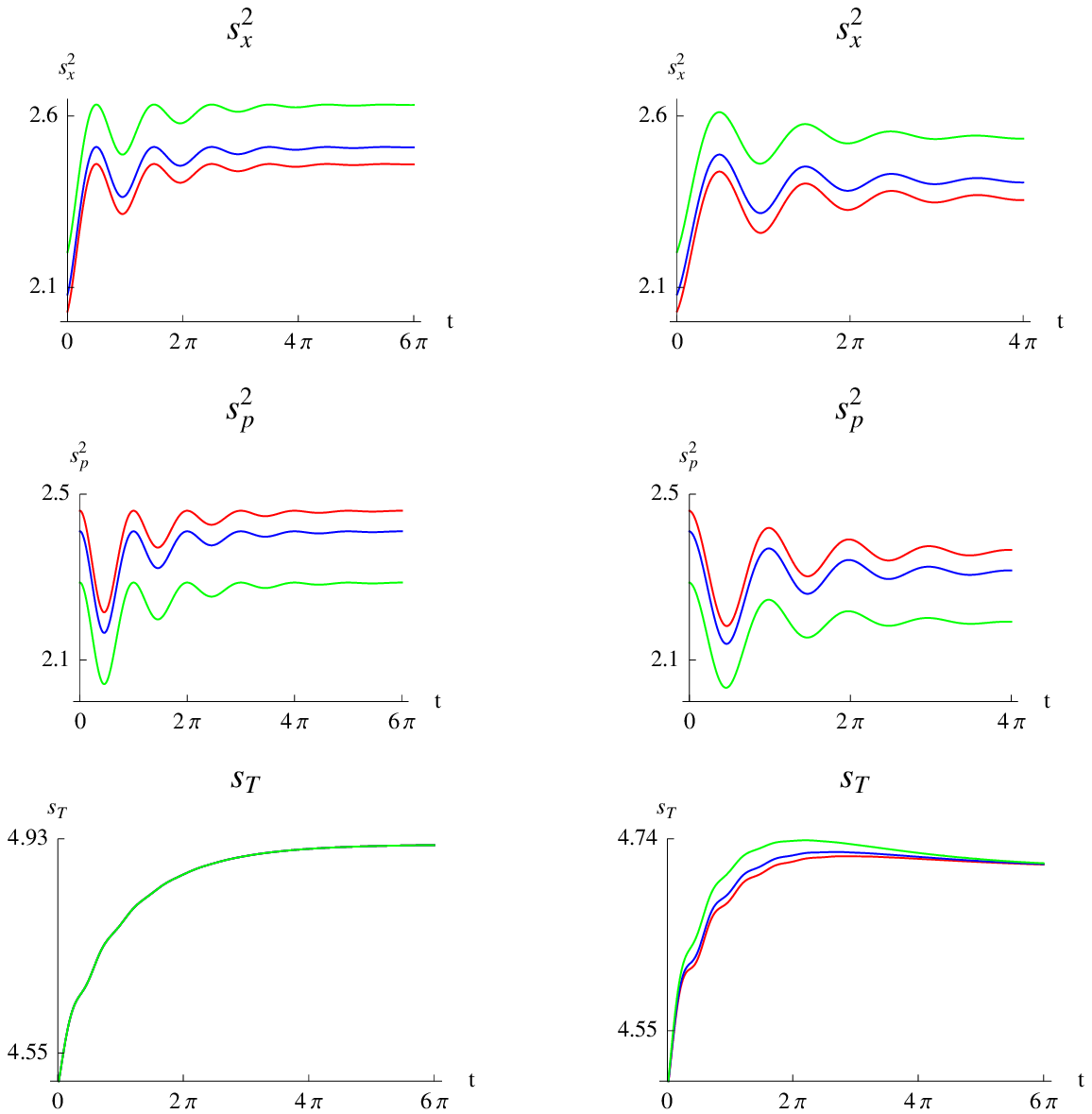}
\caption{Position space pair entropy as a function of time, $s_{x}^{2}(t)$, Eq. (\ref{sx2}) (top), momentum space pair entropy as a function of time, $s_{p}^{2}(t)$, Eq. (\ref{sp2}) (middle) and total pair entropy as a function of time, $s_{T} (t)$ (bottom), of the Moshinsky atom, for the case of pure-dephasing without relaxation (left) and relaxation without pure-dephasing (right). $\omega=1$ and $\gamma =0.15$. $\lambda = 0$ (red), $\lambda = 0.3$ (blue), $\lambda = 0.5$ (green).}
\label{fig6}
\end{center}
\end{figure}

\begin{figure}[H]
\begin{center}
\includegraphics{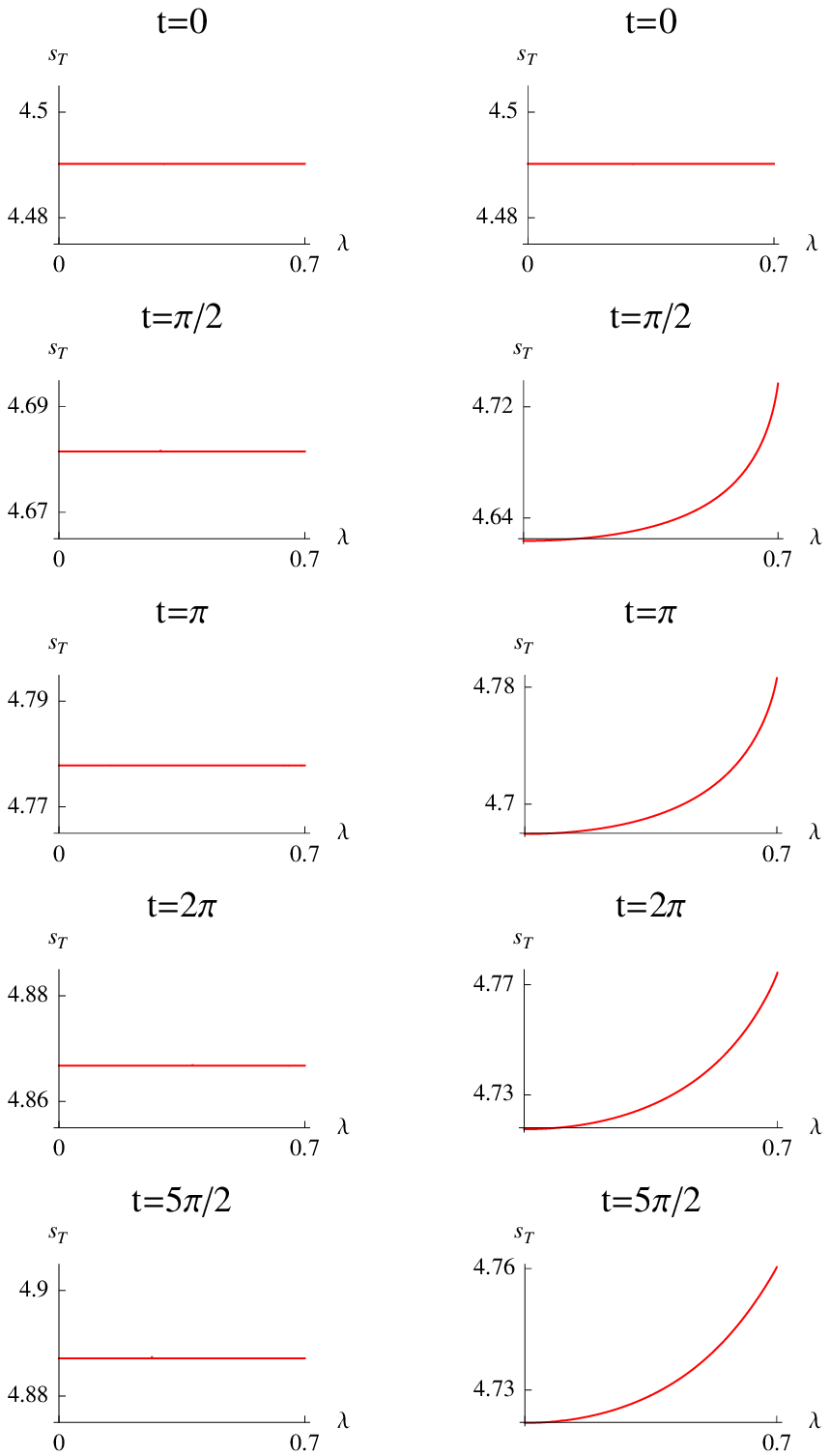}
\caption{Total pair entropy as a function of the interparticle potential, $s_{T} (\lambda)$, of the Moshinsky atom, for the case of pure-dephasing without relaxation (left) and relaxation without pure-dephasing (right). $\omega=1$ and $\gamma =0.15$, for different values of time, $t$.}
\label{fig7}
\end{center}
\end{figure}

\begin{figure}[H]
\begin{center}
\includegraphics{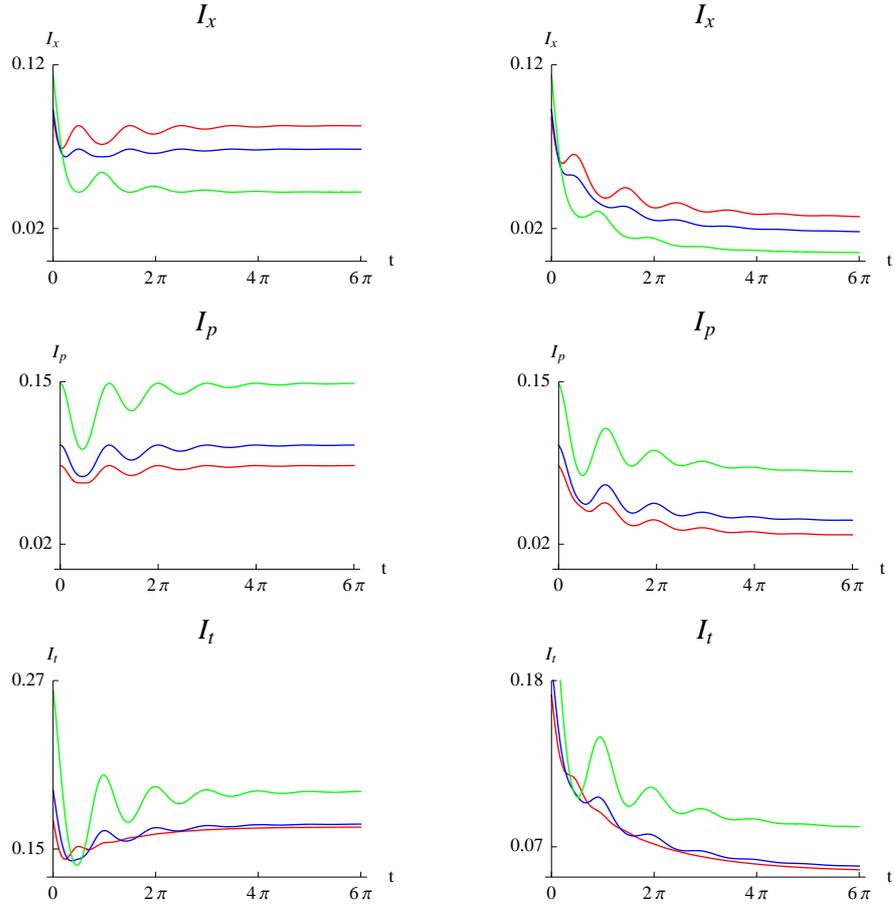}
\caption{Position space mutual information as a function of time, $I_{x}(t)$, Eq. (\ref{mutinfx}) (top), momentum space mutual information as a function of time, $I_{p}(t)$, Eq. (\ref{mutinfp}) (middle) and total mutual information as a function of time, $I_{t} (t)$ (bottom), of the Moshinsky atom, for the case of pure-dephasing without relaxation (left) and relaxation without pure-dephasing (right). $\omega=1$ and $\gamma =0.15$. $\lambda = 0$ (red), $\lambda = 0.3$ (blue), $\lambda = 0.5$ (green).}
\label{fig8}
\end{center}
\end{figure}

\begin{figure}[H]
\begin{center}
\includegraphics{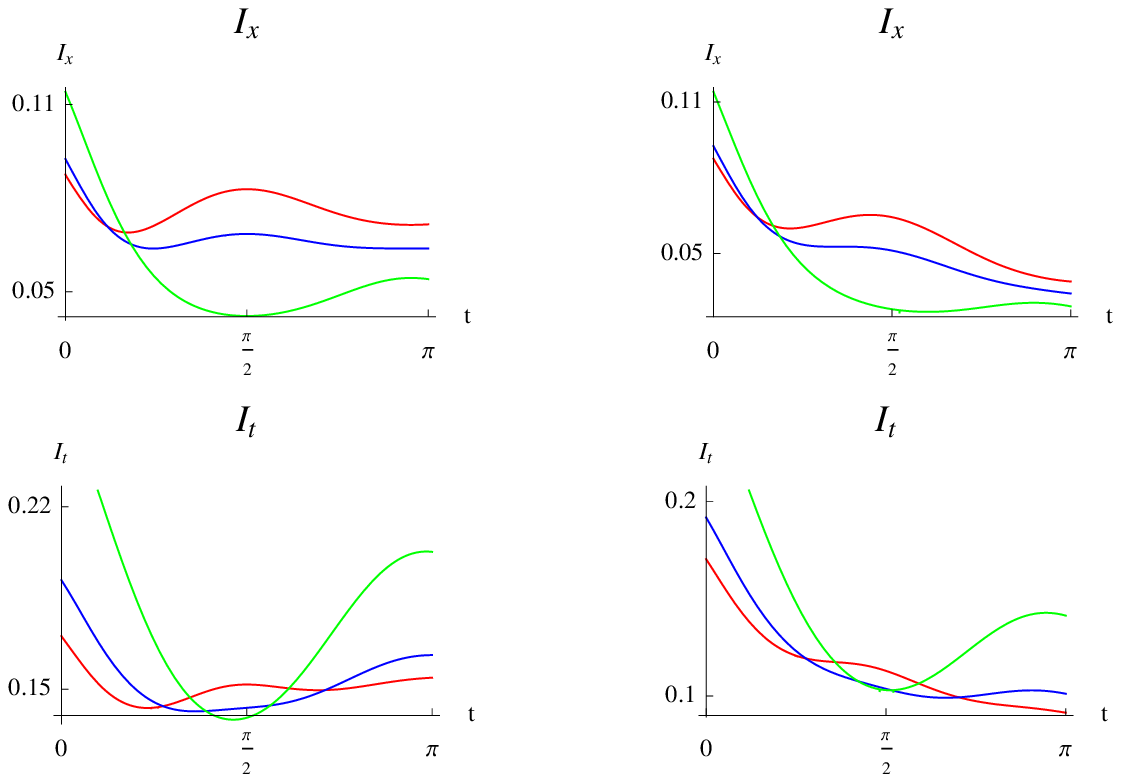}
\caption{Detail of the previous plot. Position space mutual information as a function of time, $I_{x}(t)$, Eq. (\ref{mutinfx}) (top) and total mutual information as a function of time, $I_{t} (t)$ (bottom), of the Moshinsky atom, for the case of pure-dephasing without relaxation (left) and relaxation without pure-dephasing (right). $\omega=1$ and $\gamma =0.15$. $\lambda = 0$ (red), $\lambda = 0.3$ (blue), $\lambda = 0.5$ (green).}
\label{fig9}
\end{center}
\end{figure}

\begin{figure}[H]
\begin{center}
\includegraphics{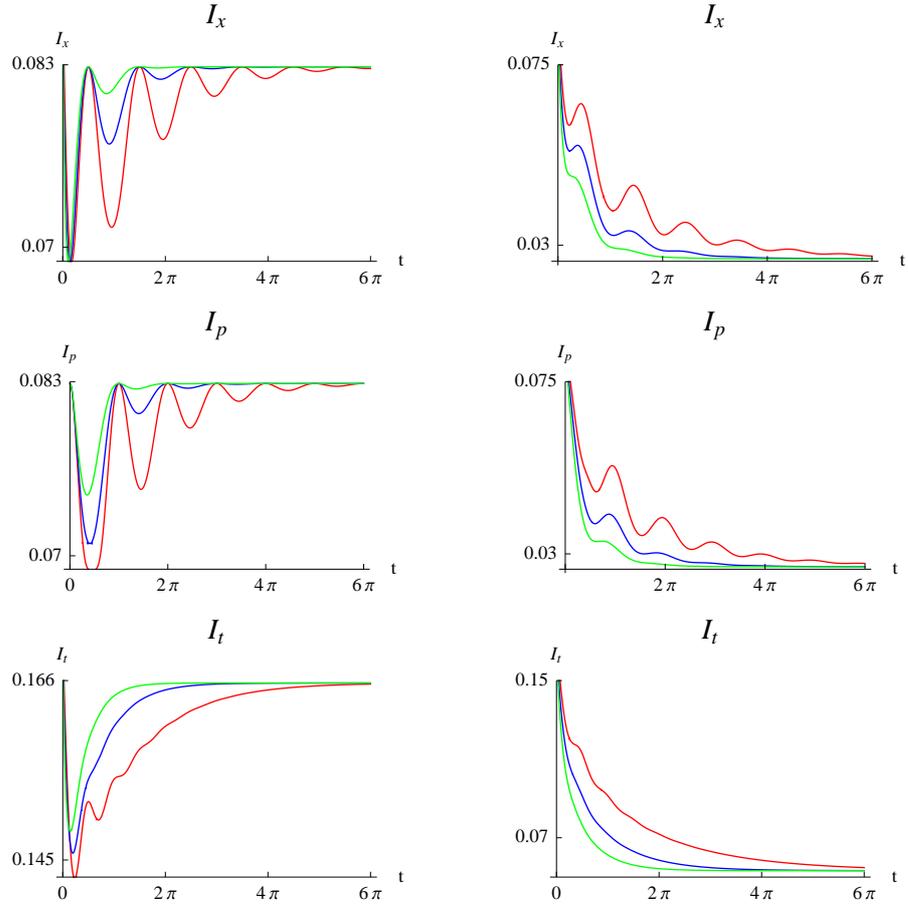}
\caption{Position space mutual information as a function of time, $I_{x}(t)$, Eq. (\ref{mutinfx}) (top), momentum space mutual information as a function of time, $I_{p}(t)$, Eq. (\ref{mutinfp}) (middle) and total mutual information as a function of time, $I_{t} (t)$ (bottom), for the case of pure-dephasing without relaxation (left) and relaxation without pure-dephasing (right) in the non-interacting Moshinsky atom ($\lambda =0$). $\omega=1$ and $\gamma = 0.15$ (red), $\gamma = 0.3$ (blue), $\gamma = 0.5$ (green).}
\label{fig10}
\end{center}
\end{figure}

\begin{figure}[H]
\begin{center}
\includegraphics{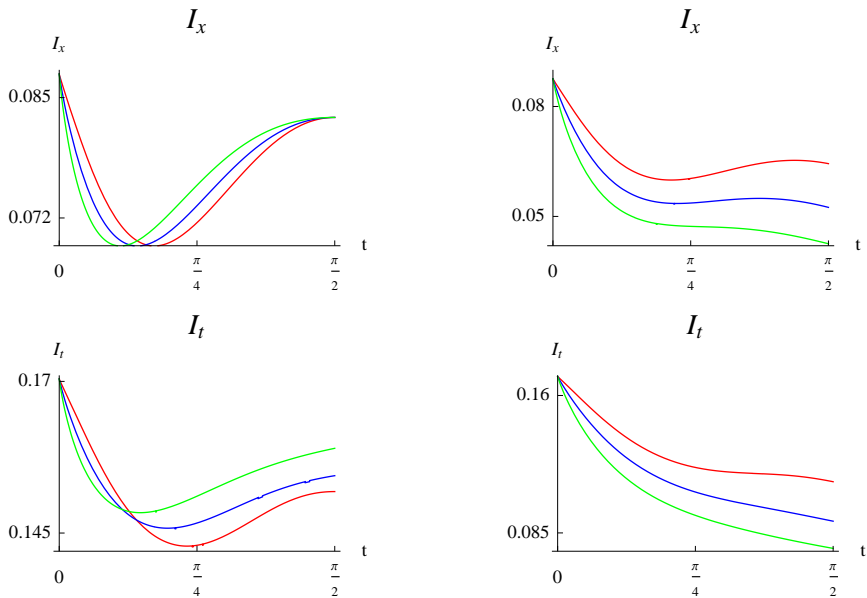}
\caption{Detail of the previous plot. Position space mutual information as a function of time, $I_{x}(t)$, Eq. (\ref{mutinfx}) (top) and total mutual information as a function of time, $I_{t} (t)$ (bottom), for the case of pure-dephasing without relaxation (left) and relaxation without pure-dephasing (right) in the non-interacting Moshinsky atom ($\lambda =0$). $\omega=1$ and $\gamma = 0.15$ (red), $\gamma = 0.3$ (blue), $\gamma = 0.5$ (green).}
\label{fig11}
\end{center}
\end{figure}

\begin{figure}[H]
\begin{center}
\includegraphics{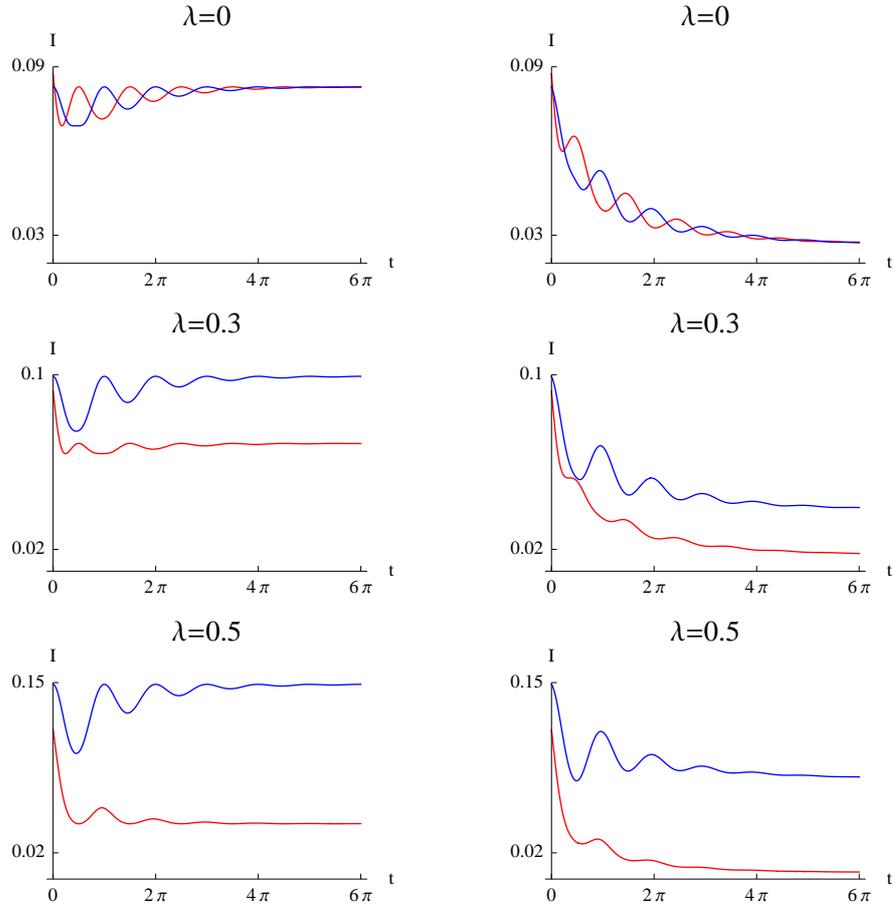}
\caption{Comparison between position space mutual information as a function of time, $I_{x}(t)$, Eq. (\ref{mutinfx}) (red) and momentum space mutual information as a function of time, $I_{p}(t)$, Eq. (\ref{mutinfp}) (blue), of the Moshinsky atom, for the case of pure-dephasing without relaxation (left) and relaxation without pure-dephasing (right). $\omega=1$ and $\gamma =0.15$. $\lambda = 0$ (top), $\lambda = 0.3$ (middle), $\lambda = 0.5$ (bottom).}
\label{fig12}
\end{center}
\end{figure}


\newpage

\begin{thebibliography}{}
\bibitem{barends2014a} R. Barends, J. Kelly, A. Megrant, A. Veitia, D. Sank, E. Jeffrey, T.C. White, J. Mutus, A.G. Fowler, B. Campbell, Y. Chen, Z. Chen, B. Chiaro, A. Dunsworth, C. Neill, P. O'Malley, P. Roushan, A. Vainsencher, J. Wenner, A.N. Korotkov, A.N. Cleland and J.M. Martinis, Nature, {\bf 508}, 300 (2014).
\bibitem{fowler2014} A.G. Fowler, D. Sank, J. Kelly, R. Barends, J.M. Martinis, arXiv:1405.1454v1 (2014).
\bibitem{barends2014b} R. Barends, J. Kelly, A. Veitia, A. Megrant, A.G. Fowler, B. Campbell, Y. Chen, Z. Chen, B. Chiaro, A. Dunsworth, I.-C. Hoi, E. Jeffrey, C. Neill, P.J.J. O'Malley, J. Mutus, C. Quintana, P. Roushan, D. Sank, J. Wenner, T.C. White, A.N. Korotkov, A.N. Cleland and J.M. Martinis, Phys. Rev. A, {\bf 90}, 030303(R) (2014).
\bibitem{kelly2015} J. Kelly, R. Barends, A.G. Fowler, A. Megrant, E. Jeffrey, T.C. White, D. Sank, J.Y. Mutus, B. Campbell, Y. Chen, Z. Chen, B. Chiaro, A. Dunsworth, I.-C. Hoi, C. Neill, P.J.J. O'Malley, C. Quintana, P. Roushan, A. Vainsencher, J. Wenner, A.N. Cleland and J.M. Martinis, Nature, {\bf 519}, 66 (2015).
\bibitem{corcoles2015} A.D. C{\'o}rcoles, E. Magesan, S.J. Srinivasan, A.W. Cross, M. Steffen, J.M. Gambetta and J.M. Chow, Nature Comm., {\bf 6}, 6979 (2015).
\bibitem{caldeira1985} A.O. Caldeira and A.J. Leggett, Phys. Rev. A, {\bf 31}, 1059 (1985).
\bibitem{bedingham2013} D.J. Bedingham and J.J. Halliwell, Phys. Rev. A, {\bf 88}, 022128 (2013).
\bibitem{hu1992} B.L. Hu, J.P. Paz, and Y. Zhang, Phys. Rev. D, {\bf 45}, 2843 (1992).
\bibitem{valdes2015} A. Vald{\'e}s-Hern{\'a}ndez, A.P. Majtey and A.R. Plastino, Phys. Rev. A, {\bf 91}, 032313 (2015).
\bibitem{chou2008a} C.-H. Chou, T. Yu and B.L. Hu, Phys. Rev. E, {\bf 77}, 011112 (2008).
\bibitem{chou2008b} C.-H. Chou, B.L. Hu and T. Yu, Phys. A, {\bf 387}, 432 (2008).
\bibitem{abe1990} S. Abe and N. Suzuki, Phys. Rev. A, {\bf 41}, 4608 (1990).
\bibitem{anderson1993} A. Anderson and J.J. Halliwell, Phys. Rev. D, {\bf 48}, 2753 (1993).
\bibitem{hu1993} B.L. Hu and Y. Zhang, Mod. Phys. Lett. A, {\bf 8}, 3575 (1993).
\bibitem{hu1995} B.L. Hu and Y. Zhang, Int. J. Mod. Phys. A, {\bf 10}, 4537 (1995).
\bibitem{ponomarenko2001} S.A. Ponomarenko and E. Wolf, Phys. Rev. A, {\bf 63}, 062106 (2001).
\bibitem{isar2002} A. Isar and W. Scheid, Phys. Rev. A, {\bf 66}, 042117 (2002).
\bibitem{schroeder2006} M. Schr{\"o}eder, U. Kleinekath{\"o}fer and M. Schreiber, J. Chem. Phys., {\bf 124}, 084903 (2006).
\bibitem{tretiak2000} S. Tretiak, C. Middleton, V. Chernyak and S. Mukamel, J. Phys. Chem. B, {\bf 104}, 4519 (2000).
\bibitem{schroeder2007} M. Schr{\"o}eder, M. Schreiber and U. Kleinekath{\"o}fer, J. Lumin., {\bf 125}, 126 (2007).
\bibitem{ranger2010} T. Renger and R.A. Marcus, J.Chem. Phys., {\bf 116}, 9997 (2010).
\bibitem{burke2005} K. Burke, R. Car and R. Gebauer, Phys. Rev. Lett., {\bf 94}, 46803 (2005).
\bibitem{gebauer2005} R. Gebauer, S. Piccinin and R. Car, Chem. Phys. Chem., {\bf 6}, 1727 (2005).
\bibitem{gebauer2004} R. Gebauer and R. Car, Phys. Rev. Lett., {\bf 93}, 160404 (2004).
\bibitem{zheng2010} X. Zheng, G. Chen, Y. Mo, S. Koo, H. Tian, C. Yam and Y. Yan, J. Chem. Phys., {\bf 133}, 114101 (2010).
\bibitem{zheng2007} X. Zheng, F. Wang, C. Yam, Y. Mo and G. Chen, Phys. Rev. B, {\bf 75}, 195127 (2007).
\bibitem{kurth2005} S. Kurth, G. Stefanucci, C.-O. Ambladh, A. Rubio and E.K.U. Gross, Phys. Rev. B, {\bf 72}, 035308 (2005).
\bibitem{perdomo2010} A. Perdomo, L. Vogt, A. Najmaie and A. Aspuru-Guzik, Appl. Phys. Lett., {\bf 96}, 093114 (2010).
\bibitem{rebentrost2010} P. Rebentrost, M. Stopa and A. Aspuru-Guzik, Nano Lett., {\bf 10}, 2849 (2010).
\bibitem{rebentrost2009} P. Rebentrost, R. Chakraborty and A. Aspuru-Guzik, J. Chem. Phys., {\bf 131}, 184102 (2009).
\bibitem{abramavicius2004} D. Abramavicius and S. Mukamel, J. Phys. Chem. B, {\bf 108}, 10295 (2004).
\bibitem{dunkel2005} J. Dunkel and S.A. Trigger, Phys. Rev. A, {\bf 71}, 052102 (2005).
\bibitem{garbaczewski2005a} P. Garbaczewski, Phys. Rev. A, {\bf 72}, 056101 (2005).
\bibitem{garbaczewski2005b} P. Garbaczewski, Entropy, {\bf 7}, 253 (2005).
\bibitem{garbaczewski2006} P. Garbaczewski, J. Stat. Phys., {\bf 123}, 315 (2006).
\bibitem{haldar2013} S.K. Haldar and B. Chakrabarti, Int. J. Mod. Phys. B, {\bf 27}, 1350048 (2013).
\bibitem{aguiar2014} V. Aguiar, I. Guedes, Phys. A, {\bf 401}, 159 (2014).
\bibitem{mosha} M. Moshinsky, Am. J. Phys., {\bf 36}, 52 (1968).
\bibitem{moshb} M. Moshinsky, Am. J. Phys., {\bf 36}, 763 (1968).
\bibitem{shannon} C.E. Shannon, Bell Syst. Tech. J., {\bf 27}, 379 (1948).
\bibitem{cover} T.M. Cover and J.A. Thomas, Elements of Information Theory, John Wiley and Sons, New York, 1991.
\bibitem{bbm} I. Bialynicki-Birula and J. Mycielski, Commun. Math. Phys., {\bf 44}, 129 (1975).
\bibitem{guevarajcp} N.L. Guevara, R.P. Sagar and R.O. Esquivel, J. Chem. Phys., {\bf 119}, 7030 (2003).
\bibitem{sagar2005} R.P. Sagar and N.L. Guevara, J. Chem. Phys., {\bf 123}, 044108 (2005).
\bibitem{sagar2006} R.P. Sagar and N.L. Guevara, J. Chem. Phys., {\bf 124}, 134101 (2006).
\bibitem{tempel2011} D.G. Tempel and A. Aspuru-Guzik, Chem. Phys., {\bf 391}, 130 (2011).
\bibitem{oqs1} H.P. Breuer, F. Petruccione, The Theory of Open Quantum Systems, Oxford
University Press, USA, 2002.
\bibitem{yuen2010} J. Yuen-Zhou, D.G. Tempel, C.A. Rodr{\'i}guez-Rosario and A. Aspuru-Guzik, Phys. Rev. Lett., {\bf 104}, 043001 (2010).
\bibitem{tempel2012} D.G. Tempel, J. Yuen-Zhou and A. Aspuru-Guzik, in: M.A.L. Marques, N.T. Maitra, F.M.S. Nogueira, E.K.U. Gross and A. Rubio (Eds.), Fundamentals of Time-Dependent Density Functional Theory, 837. Berlin, Heidelberg: Springer Berlin Heidelberg, 2012, pp. 211-229.
\bibitem{yuen2013} J. Yuen-Zhou and A. Aspuru-Guzik, Phys. Chem. Chem. Phys., {\bf 15}, 12626 (2013).
\bibitem{yuen2009} J. Yuen-Zhou, C. Rodr{\'i}guez-Rosario and A. Aspuru-Guzik, Phys. Chem. Chem. Phys., {\bf 11}, 4509 (2009).
\bibitem{guevara} N.L. Guevara, R.P. Sagar and R.O. Esquivel, Phys. Rev. A, {\bf 67}, 012507 (2003).
\bibitem{lagunapra} H.G. Laguna and R. P. Sagar, Phys. Rev. A, {\bf 84}, 012502 (2011).
\end{thebibliography}
\end{document}